# Research on Interpore Distance of Anodic Aluminum Oxide Template

**Xue-jie Liu, Liang-fang Li ***

*School of Mechanical Engineering, Inner Mongolia University of Science & Technology, Baotou, Inner Mongolia 014010, PR China. E-mail: liliangfangllf@163.com*

**Abstract**

The relationship between the interpore of anodic aluminum oxide (AAO) template and the influencing factors of electrolyte, temperature and oxidation voltage etc. was researched and summarized in this paper. It was pointed out that the interpore was influenced mostly by electrolyte type and oxidation voltage, and least by the electrolyte concentration and oxidation temperature. The interpore of AAO template increases with the oxidation voltage increases. By adjusting the electrolyte and oxidation voltage, a desired interpore of template can be acquired. To acquire a large interpore template can use electrolyte of phosphoric acid or chromic acid under a comparatively higher oxidation voltage. To acquire a small interpore template (with interpore within 10nm) can use sulfate electrolyte under a comparatively lower oxidation voltage or pulse voltage. Alternatively, oxidizing with lower oxidation voltage in a mixed electrolyte of $H_2SO_4$ and $Al_2$ $(SO4)_3$, whereafter immersion in the mixture of HCl and $CuCl_2$ to corrode the template for some time, until the angles of cell appear six holes with small pore diameter, the interpore of AAO template decrease to about 0.6 times of the original.

## 1 Introduction

Since Keller et al [1] has reported that using electrochemical method prepared porous alumina templates (PAT) in 1953, preparing PAT with highly ordered hole has been the focus of researchers' study. In 1995, Masuda et al [2] reported preparing highly ordered Porous anodic alumina (PAA) template used twice anodic oxidation method and preparing metal nanowires used PAA template in Science. Since then, with the PAA template scientists have prepared a variety of nano-structure material successfully [3].

The grain size of Nano structure material is from few to dozens nanometer which brings about a series of different physical and mechanical properties [4-13]. The porous anodic alumina template has been widely used in preparation of nanometer structure materials with its excellent characteristics. However, as an important template which has interposed distance between decades and hundred nanometer [14-17], it not reaches the requirement of preparation of Nano structure material with size in few nanometers [18].

Masuda et al [19] studied the effect of surface condition on the ordering of pores. Shingubara, Keller and Jamginas et al [1, 20, 21] reported that the interpore distance of AAO template increase with anodizing voltage with approximate linear relationship. It has been studied more extensively that the direct relationship between the interpore of AAO template and anode voltage in the subsequent research [22-43]. Due to the strict requirement of the size of nano structure material, it is the key point that preparation of the interpore distance meets the requirement of anodic alumina template. It studied the relationship between the interpore of anodic aluminum oxide (AAO) template and the influencing factors of electrolyte, temperature, especially oxidation voltage to provide reference for particular interpore template.

## 2 Influence factors for interpore distance

### 2.1 Electrolyte

Electrolyte is the key factor，which effects the interpore distance of porous anodic alumina template including electrolyte type and electrolyte concentration. Tab.1 shows that Wojciech J. S., and J. Gong etc. adopted the same concentration as the electrolyte [36, 40]. When the oxidizing voltage is 10 V, the interpore is 30 nm. When oxidizing voltage is 25 V, the interpore is 58.9 nm. Y. C. Zhao and K. H. Lee et al [14,15,36,41] study the same concentration oxalic acid electrolyte. In the research, when the oxidizing voltage is from 25 V to 170 V, the interpore is from 91.9nm to 371 nm. Jaime M. and Masuda H. et al [37, 44] adopted the same concentration as the electrolyte. In the result, when the oxidizing voltage is from 185 V to 195 V, the interpore is from 450nm to 500 nm. And then Leszke Z. et al [39] used chromic acid as the electrolyte. In the research, when the oxidizing voltage is 40 V and 50 V, the interpore is 136.8nm and 171 nm.

S. Y. Zhao and X. W. Fu et al [38, 43] reported that sulfuric acid with different concentration were used for electrolyte, and under the other same conditions, the size of alumina template hole spacing was increasing with the reduce of the concentration of electrolyte, but the scale of influence is small, it was shown in table 2.

**2.2 Temperature**

Leszek Z. et al [39] reported, under the other same conditions, chromic acid was used for electrolyte, and alumina template was made by changing the temperature. The result showed that with the increasing of the temperature, the interpore of anodic aluminum oxide template will increase too, but the increasing range is small. The research from K. H. Lee and W. Lee et al [15, 16, 22, 24, 27, 36-38] showed that the interpore of template increased with temperature too, as it is shown in table 3.

**2.3 Voltage**

The influence of anode oxidation voltage concerning the interpore of template has been researched proverbially, in a certain oxidation voltage range, the interpore of template with voltage synchronous increase has been widely argued. F. zhang and S. Y. Zhao et al [28, 38] used sulfuric acid with concentration 10wt% for electrolyte, oxidation voltage between 6-18V, the interpore of template from 20.5 to 58.3 nm. In the range of voltage from 6 to 18 V, it has a linear relationship between voltage and the interpore. K. M. Song and Wojciech J. S. et al [36, 37, 40] used 0.3M sulfuric acid for electrolyte, the research result is the same to the above, the range of the voltage is from 10V to 32 V, and the interpore is from 30nm to 114 nm. The proportional relation is not only suitable for sulfuric acid electrolyte, but also suitable for oxalic acid electrolyte. The report from Y. C. Zhao and J. Gong et al [14, 40-42], 0.3M oxalic acid was used for electrolyte, under the condition of anode voltage from 15V to 170V, the interpore of template is between 40nm and 371 nm. In the research of Jaime M. and Masuda H. et al [37, 44] who used phosphoric acid for electrolyte, is also suitable for the relationship. As it is shown in table4. Leszek Z. et al [39] used chromic acid as electrolyte, the oxidation voltage was from 20V to 50V and the interpore o was from 68.4nm to171nm.

Above the researchers' studies, their research results followed the positively proportional relationship between the interpore of template and the anode voltage. But this proportion will be breached if the certain range of oxidation voltage is exceeded [26]. The proportional relationship varies with different types of the electrolyte.

In addition, the interpore of AAO template is influenced when it use different oxidation methods. First imprinting on the surface of aluminum and then oxidizing, J. Choi and Namyong K et al [28, 44] could get template with larger interpore and highly orderly degree. S. Y. Zhao et al [38] used the mixture acid with 0.8 M $H_2SO_4$ and 0.1M $Al_2(SO4)_3$ as electrolyte. After the first step oxidation, the pore diameter is 55 nm and the interpore is 85 nm. And then using the mixture with 19% HCl and 0.2M $CuCl_2$, it can remove residual aluminum. After soaked for 40 minutes, the original pores membrane is blocked, but small pores which pore diameter is 15 nm appears at between the places of the six angles Hexagonal cell. After soaked for one hour, the pore diameter might increase to 25nm. After the closing pores were opened, the surface of alumina membrane appeared pores that the small pores around the big pores. This greatly reduced the interpore distance, the existing interpore was about 0.6 times of the original.

**3 Conclusions**

From series of summaries, we realized that the interpore of AAO template has intimate connections to the electrolyte ingredient, electrolyte concentration, temperature and oxidation voltage. What the difference is, compare with the influence from electrolyte ingredient and oxidation voltage, although the electrolyte concentration and oxidation temperature has influence to the interpore, they can be ignored. On condition that all other qualifications are the same, use sulphuric acid, oxalic acid, phosphoric acid, and chromic acid as electrolyte, the corresponding the interpore increase successively. On condition that all other qualifications are the same, use a different oxidation voltage which is controlled in a certain scope, the result is that the interpore of AAO template increase complies with the increase of the voltage. According to the different requirements of the interpore of template, needed template can be fabricated by adjusting the electrolyte and oxidation voltage. Big interpore of template fabrication could use phosphoric acid or chromic acid as electrolyte with high oxidation voltage. In order to acquire small interpore of template (the interpore < 20nm) could use sulphuric acid as electrolyte with low (< 6V) oxidation voltage, but the oxidation reaction can't work with small voltage, with the help of higher pulse voltage the problem will be solved. Alternatively, use $H_2SO_4$ and $Al_2$ $(SO4)_3$ as mixed electrolyte. After oxidized, immersion in the HCl and CuCl2 mixed electrolyte, on condition that the holes are not breakthrough, make the small holes diameter close to the original big holes diameter, the hole-hole spaces could decrease to 0.6 times of the original holes-holes spaces. This information has a significant reference value for us to fabricating anodic aluminum oxide templates with small interpore about 10nm.

**Acknowledgment**

This research was supported by the National Natural Science Foundation of China (NNSFC) under grant 50845065 and Inner Mongolia Natural Science Foundation under grant 2010Zd21.

Table 1. The relationship between the types of electrolytes and the interpore distance of anodic aluminum oxide template

| First step anodization condition | | | | | Second step anodization condition | | | | | pore diameter (nm) | Interpore spacing (nm) | References |
|---|---|---|---|---|---|---|---|---|---|---|---|---|
| Electrolyte | Electrolyte Concentration (mol/L) | Voltage(V) | Temperature(°C) | Time(h) | Electrolyte | Electrolyte Concentration (mol/L) | Voltage(V) | Temperature(°C) | Time(h) | | | |
| $H_2SO_4$ | 0.3 | 10 | 3 | 7 | $H_2SO_4$ | 0.3 | 10 | 3 | 2 | 20 | 30 | [40] |
| $H_2SO_4$ | 0.3 | 25 | 0 | 4 | $H_2SO_4$ | 0.3 | 25 | 0 | 4 | 17.9 | 58.9 | [36] |
| Oxalic | 0.3 | 25 | 0 | 4 | Oxalic | 0.3 | 25 | 0 | 4 | 21.5 | 91.9 | [36] |
| Oxalic | 0.3 | 40 | 0 | 3 | Oxalic | 0.3 | 40 | 0 | 3 | 30 | 100 | [15] |
| Oxalic | 0.3 | 50-60 | 0 | 3 | Oxalic | 0.3 | 50-60 | 0 | 8 | 76-80 | 130-150 | [14] |
| Oxalic | 0.3 | 115-170 | 0 | 12 | Oxalic | 0.3 | 115-170 | 0 | 24 | - | 257-371 | [41] |
| $H_3PO_4$ | 0.3 | 185 | 3 | - | $H_3PO_4$ | 0.3 | 185 | 3 | - | - | 450 | [37] |
| $H_3PO_4$ | 0.3 | 195 | 0 | 1 | - | - | - | - | - | - | 500 | [44] |
| chromic | 0.3 | 40 | 20 | 1 | chromic | 0.3 | 40 | 20 | 1 | 28.4 | 136.8 | [39] |
| chromic | 0.3 | 50 | 20 | 1 | chromic | 0.3 | 50 | 20 | 1 | 35.5 | 171 | [39] |

Table 2. The relationship between electrolyte concentrations and the interpore distance of anodic aluminum oxide template

| First step anodization condition | | | | | Second step anodization condition | | | | | pore diameter (nm) | Interpore spacing (nm) | References |
|---|---|---|---|---|---|---|---|---|---|---|---|---|
| Electrolyte | Electrolyte Concentration (mol/L) | Voltage(V) | Temperature(°C) | Time(h) | Electrolyte | Electrolyte Concentration (mol/L) | Voltage(V) | Temperature(°C) | Time(h) | | | |
| $H_2SO_4$ | 0.5 | 20 | 0 | 2 | $H_2SO_4$ | 0.5 | 20 | 0 | 2 | 35 | 57 | [43] |
| $H_2SO_4$ | 0.6 | 20 | 0 | 2 | $H_2SO_4$ | 0.6 | 20 | 0 | 2 | 30 | 52 | [43] |
| $H_2SO_4$ | 0.8 | 20 | 0 | 2 | $H_2SO_4$ | 0.8 | 20 | 0 | 2 | 25 | 47 | [43] |
| $H_2SO_4$ | 3 wt% | 12.5 | 0 | 1 | $H_2SO_4$ | 3 wt% | 12.5 | 0 | 24 | 19.9 | 32.7 | [38] |
| $H_2SO_4$ | 5wt% | 12.5 | 0 | 1 | $H_2SO_4$ | 3 wt% | 12.5 | 0 | 24 | 19.1 | 31.9 | [38] |
| $H_2SO_4$ | 7 wt% | 12.5 | 0 | 1 | $H_2SO_4$ | 3 wt% | 12.5 | 0 | 24 | 19 | 31.8 | [38] |
| $H_2SO_4$ | 10 wt% | 12.5 | 0 | 1 | $H_2SO_4$ | 3 wt% | 12.5 | 0 | 24 | 18.4 | 32.6 | [38] |
| $H_2SO_4$ | 20 wt% | 12.5 | 0 | 1 | $H_2SO_4$ | 3 wt% | 12.5 | 0 | 24 | 13.7 | 29.9 | [38] |

Table 3. The relationship between temperature and the interpore distance of anodic aluminum oxide template

| First step anodization condition | | | | | Second step anodization condition | | | | | pore diameter (nm) | Interpore spacing (nm) | References |
|---|---|---|---|---|---|---|---|---|---|---|---|---|
| Electrolyte | Electrolyte Concentration (mol/L) | Voltage(V) | Temperature(°C) | Time(h) | Electrolyte | Electrolyte Concentration (mol/L) | Voltage(V) | Temperature(°C) | Time(h) | | | |
| $H_2SO_4$ | 10 wt% | 12.5 | 0 | 1 | $H_2SO_4$ | 3 wt% | 12.5 | 0 | 24 | 18.4 | 32.6 | [38] |
| $H_2SO_4$ | 10wt% | 12 | 3 | 12 | $H_2SO_4$ | 10wt% | 12 | 3 | 3 | 11.1 | 33.5 | [22] |
| $H_2SO_4$ | 0.3 | 25 | 0 | 4 | $H_2SO_4$ | 0.3 | 25 | 0 | 4 | 17.9 | 58.9 | [36] |
| $H_2SO_4$ | 0.3 | 27-32 | 1 | 3 | - | - | - | - | - | 15-30 | 78-114 | [24] |
| $H_2SO_4$ | 0.3 | 26 | 3 | - | $H_2SO_4$ | 0.3 | 26 | 3 | - | - | 60 | [37] |
| Oxalic | 0.3 | 40 | 0 | 3 | Oxalic | 0.3 | 40 | 0 | 3 | 30 | 100 | [15] |
| Oxalic | 0.3 | 40 | 1 | 2.6 | - | - | - | - | - | 40 | 100 | [16] |
| Oxalic | 0.3 | 40 | 4 | 0.5 | Oxalic | 0.3 | 40 | 4 | 1 | 45 | 100 | [39] |
| Oxalic | 0.3 | 40 | 15 | - | Oxalic | 0.3 | 40 | 15 | - | - | 108 | [37] |
| Oxalic | 0.3 | 40 | 35 | 0.5 | Oxalic | 0.3 | 40 | 35 | 0.5 | 47 | 104 | [27] |
| chromic | 0.3 | 20 | 20 | 1 | chromic | 0.3 | 20 | 20 | 1 | 14.2 | 68.4 | [39] |
| chromic | 0.3 | 20 | 30 | 1 | chromic | 0.3 | 20 | 30 | 1 | 18.7 | 68.6 | [39] |
| chromic | 0.3 | 20 | 40 | 1 | chromic | 0.3 | 20 | 40 | 1 | 19.3 | 71.8 | [39] |

Table 4. The relationship between anodizing voltage and the interpore distance of anodic aluminum oxide template

| First step anodization condition | | | | | Second step anodization condition | | | | | pore diameter (nm) | Interpore spacing (nm) | References |
|---|---|---|---|---|---|---|---|---|---|---|---|---|
| Electrolyte | Electrolyte Concentration (mol/L) | Voltage(V) | Temperature(°C) | Time(h) | Electrolyte | Electrolyte Concentration (mol/L) | Voltage(V) | Temperature(°C) | Time(h) | | | |
| $H_2SO_4$ | 10wt% | 6 | 3 | 12 | $H_2SO_4$ | 10wt% | 6 | 3 | 3 | 6.8 | 20.5 | [22] |
| $H_2SO_4$ | 10wt% | 8 | 3 | 12 | $H_2SO_4$ | 10wt% | 8 | 3 | 3 | 8.6 | 25.9 | [22] |
| $H_2SO_4$ | 10 wt% | 12.5 | 0 | 1 | $H_2SO_4$ | 3 wt% | 12.5 | 0 | 24 | 18.4 | 32.6 | [38] |
| $H_2SO_4$ | 10wt% | 12 | 3 | 12 | $H_2SO_4$ | 10wt% | 12 | 3 | 3 | 11.1 | 33.5 | [22] |
| $H_2SO_4$ | 10wt% | 16 | 3 | 12 | $H_2SO_4$ | 10wt% | 16 | 3 | 3 | 16.4 | 49.3 | [22] |
| $H_2SO_4$ | 10wt% | 18 | 3 | 12 | $H_2SO_4$ | 10wt% | 18 | 3 | 3 | 19.3 | 58.3 | [22] |
| $H_2SO_4$ | 0.3 | 10 | 3 | 7 | $H_2SO_4$ | 0.3 | 10 | 3 | 2 | 20 | 30 | [40] |
| $H_2SO_4$ | 0.3 | 25 | 0 | 4 | $H_2SO_4$ | 0.3 | 25 | 0 | 4 | 17.9 | 58.9 | [36] |
| $H_2SO_4$ | 0.3 | 26 | 3 | - | $H_2SO_4$ | 0.3 | 26 | 3 | - | - | 60 | [37] |
| $H_2SO_4$ | 0.3 | 27-32 | 1 | 3 | - | - | - | - | - | 15-30 | 78-114 | [24] |
| Oxalic | 0.3 | 15 | 5 | 3 | Oxalic | 0.3 | 15 | 5 | 2 | 30 | 40 | [40] |
| Oxalic | 0.3 | 30 | 15 | 3 | Oxalic | 0.3 | 30 | 15 | 3 | 29 | 75 | [42] |
| Oxalic | 0.3 | 40 | 15 | 3 | Oxalic | 0.3 | 40 | 15 | 3 | 37 | 96 | [42] |
| Oxalic | 0.3 | 50 | 15 | 3 | Oxalic | 0.3 | 50 | 15 | 3 | 44 | 123 | [42] |
| Oxalic | 0.3 | 60 | 0 | 3 | Oxalic | 0.3 | 60 | 0 | 8 | 80 | 150 | [14] |
| Oxalic | 0.3 | 115 | 0 | 12 | Oxalic | 0.3 | 115 | 0 | 24 | - | 257 | [41] |
| Oxalic | 0.3 | 155 | 0 | 12 | Oxalic | 0.3 | 155 | 0 | 24 | - | 346 | [41] |
| Oxalic | 0.3 | 170 | 0 | 12 | Oxalic | 0.3 | 170 | 0 | 24 | - | 371 | [41] |
| $H_3PO_4$ | 0.3 | 185 | 3 | - | $H_3PO_4$ | 0.3 | 185 | 3 | - | - | 450 | [37] |
| $H_3PO_4$ | 0.3 | 195 | 0 | 16 | - | - | - | - | - | - | 500 | [44] |
| chromic | 0.3 | 20 | 20 | 1 | chromic | 0.3 | 20 | 20 | 1 | 14.2 | 68.4 | [39] |
| chromic | 0.3 | 30 | 20 | 1 | chromic | 0.3 | 30 | 20 | 1 | 21.3 | 102.6 | [39] |
| chromic | 0.3 | 40 | 20 | 1 | chromic | 0.3 | 40 | 20 | 1 | 28.4 | 136.8 | [39] |
| chromic | 0.3 | 50 | 20 | 1 | chromic | 0.3 | 50 | 20 | 1 | 35.5 | 171 | [39] |